# Photoacoustic tracking of photo-magnetically powered nanoparticles for cancer therapy

**Jiayan Li[1], Chang Xu[2], Yingna Chen[1], Junmei Cao[1], Wanli Ye[1], Yu Cheng[2,*], Qian Cheng[1,*]**

[1] Institute of Acoustics, School of Physics Science and Engineering, Tongji University, Shanghai, 200092, China
[2] Collaborative Innovation Center for Brain Science, School of Medicine, Tongji University, Shanghai, 200092, China

[*]Email: yucheng@tongji.edu.cn, q.cheng@tongji.edu.cn

**Abstract**. The in vivo propulsion and monitoring of nanoparticles (NPs) have received tremendous achievements in the past decade. Developing functional NPs that can be efficiently manipulated inside the human body with a non-invasive tracking modality is critical to clinical translation. This study synthesized a photo-magnetically powered nanoparticle (PMN) with a $Fe_3O_4$ core and gold spiky surface. The Au-nanotips ensure PMNs have a strong light absorption in the second near-infrared (NIR) window and produce outstanding photoacoustic signals. The Bio-transmission electron microscopy and simulation results prove that the assembly of PMNs under a magnetic field further enhances the photothermal conversion in cells, contributing to the reduction of ambient viscosity. Photoacoustic imaging (PAI) realized real-time monitoring of PMN movements and revealed that laser plus magnetic coupling could improve intratumoral distribution and retention. The proposed methods exhibit excellent potential for the clinical research of cancer nanotherapies.

## 1. Introduction

Developing micro/nanoparticles has become a further emerging field for minimally invasive diagnosis and therapy of diseases [1,2]. These functional therapeutic platforms can accomplish plenty of vital tasks for biomedical applications [3], including opening cell membranes, drug delivery, and biosensing. One of the most crucial limitations that hinders the in vivo utilization of micro/nanoparticles is the efficient propulsion inside the human body aiming to break through the biological barriers [4]. To overcome these difficulties, micro/nanoparticles of versatile compositions and structures have been designed with different actuation strategies, among which the magnetic field (MF), light, and ultrasonic waves are the representative external stimuli most commonly applied [5].

Photo-magnetically powered nanoparticles (PMNs) are hybrid nanomaterials that can obtain energy from the laser and MFs [6,7]. PMNs are composed of a strong light absorption part (*e.g.*, metal or polydopamine [8], *etc*.) and a magnetic response part (*e.g.*, Fe, Ni, or iron oxide [9], *etc*.) [10]. Compared with other manipulation techniques, the MF provides a more facile strategy for the remote control of PMNs with no extra functional ligands required [11]. Furthermore, the intense light absorption in the near-infrared (NIR) window makes PMNs excellent candidates for photothermal therapy and optical IR detection [12,13].

Tracking the in vivo locomotion of micro/nanoparticles is another critical point influencing the clinical translation [14]. Imaging modality with high spatiotemporal resolution is of great interest in monitoring nanotherapeutic activities. PMNs can be detected by magnetic resonance imaging (MRI), offering outstanding contrast. However, MRI owns a temporal resolution in the millisecond range, which needs to be improved for studying various biological events [15], and demands expensive infrastructure. Other imaging methods, including ultrasound, optical coherence tomography (OCT), fluorescence, x-ray computed tomography (CT), positron emission tomography-computed tomography (PET-CT) [16], and single-photon emission computed tomography (SPECT) [17], have defects of deficient contrast, low penetration depth, and radioactive exposure. As a result, a new powerful approach that can make up for the above shortages is needed.

Photoacoustic imaging (PAI) measures the ultrasonic signals from the light absorbers irradiated by the pulsed laser. The photoacoustic (PA) tracking of medical microbots in vivo was suggested by Mariana Medina-Sánchez's group in 2017 [18]. It could reach 150-micrometer resolutions at 2–3 centimeter depths [19]. So far, the PA tracking of PMNs has been tested in brain vasculature [20], bladder, uterus [21], and tumor models [22,23] of animals, which manifests PMNs as a promising tool for clinical theranostics.

This paper devised a PMN that owned a $Fe_3O_4$ core and gold-nanotip-covered shell. We elaborated on the feasibility of using PAI to manifest the locomotion of PMNs under the administration of laser and MF coupling. The designed PMNs have a broad absorption in the NIR-II region due to the localized surface plasmon resonance (LSPR) effect and can produce strong PA signals with great photostability. Simulation works based on the Bio-transmission electron microscopy (Bio-TEM) images of cells with PMNs subjected to dual external stimuli were implemented to reveal the photothermal conversion of PMN assembly on microscales. Ex vivo PAI was conducted to test the magnetic response of PMNs. Furthermore, continuous in vivo 3D PAI was performed to track the intratumoral PMN distribution after injection.

## 2. Materials and methods

*2.1. Synthesis and characterization of photo-magnetically powered nanoparticles (PMNs)*

*2.1.1. Preparation and characterization of PMNs.* The synthesis and characterization of PMNs followed the protocols described in our previous paper [24]. The process briefly consisted of 3 steps: Firstly, a zinc-doped iron oxide (IO) core was prepared. Secondly, the gold nano-seed (GNS) was obtained based on the hydrothermal method. Thirdly, the PMNs were achieved by conjugating the GNSs to the surface of the IO. The spiky structures were formed by the GNS at last. The surface of PMN was modified to increase its cellular affinity.

Four basic properties of PMNs were studied after preparation. Firstly, the morphological structure of PMN was visualized using scanning electron microscopy (SEM) (FEI-Quanta FEG 250, America). Secondly, the size of PMN was calculated based on the transmission electron microscopy (TEM) (JEM-1230, JEOL Ltd., Japan) images. Thirdly, the metallic atom contents in the PMN were quantified utilizing the inductively coupled plasma mass spectrometry (ICP-MS) (Thermo Fisher, iCAP Q). Fourthly, an ultraviolet-visible (UV-vis) to near-infrared region (NIR) spectroscopy (Cary 90 UV-Vis, Agilent Technologies, America) was applied to determine the extinction spectrum.

The behaviors of PMNs in the bio-environment under external laser and MF were confirmed using Bio-TEM. The PMN solution (100 μg/mL, 1mL) was co-cultured with human triple-negative breast cancer cells (cell line: MDA-MB-231, $10^5$ cells/well) in a dish for 24 h, and then stimulated by the rotating magnetic field (RMF, 260 mT, 10Hz) or RMF with continuous wave (CW) laser (0.5 W/cm$^2$, 1064 nm) for 5 mins.

*2.1.2. In vitro photoacoustic (PA) signal measurements.* Figure 1(a) exhibited the experimental setup used for measuring the PA signals of PMNs in vitro. An optical parametric oscillator (Phocus Mobile, OPOTEK, Carlsbad, CA) controlled by the computer program sent a pulsed laser of 1200 nm with a 10 Hz repetition rate and 2-5 ns pulse width. The laser beam was converged via a lens into a spot of 7 mm diameter owning a fluence of 20 mJ/cm$^2$, which illuminated the PMN solution (1, 0.5, 0.25, 0.125, 0.0625 g/mL, 0.8 mL) through the opening of a glass tube. The tube was placed on a glass brick in the water tank to ensure the opening slightly (~3 mm) beyond the water surface. A needle hydrophone (HNC-1500, ONDA, Sunnyvale, CA) was exploited to collect the PA signals stemming from the material solution. The hydrophone was placed at 20° to the horizon and pointed at the center of the tube. The PA signals were amplified by 35 dB via an amplifier (5072PR, OLYMPUS, Tokyo, Japan) and stored by an oscilloscope (HDO6000, Teledyne Leroy, USA) at a sampling rate of 2500 MHz and 64 average times. A continuous PA measurement for 60 seconds was performed to test the photostability of PMNs.

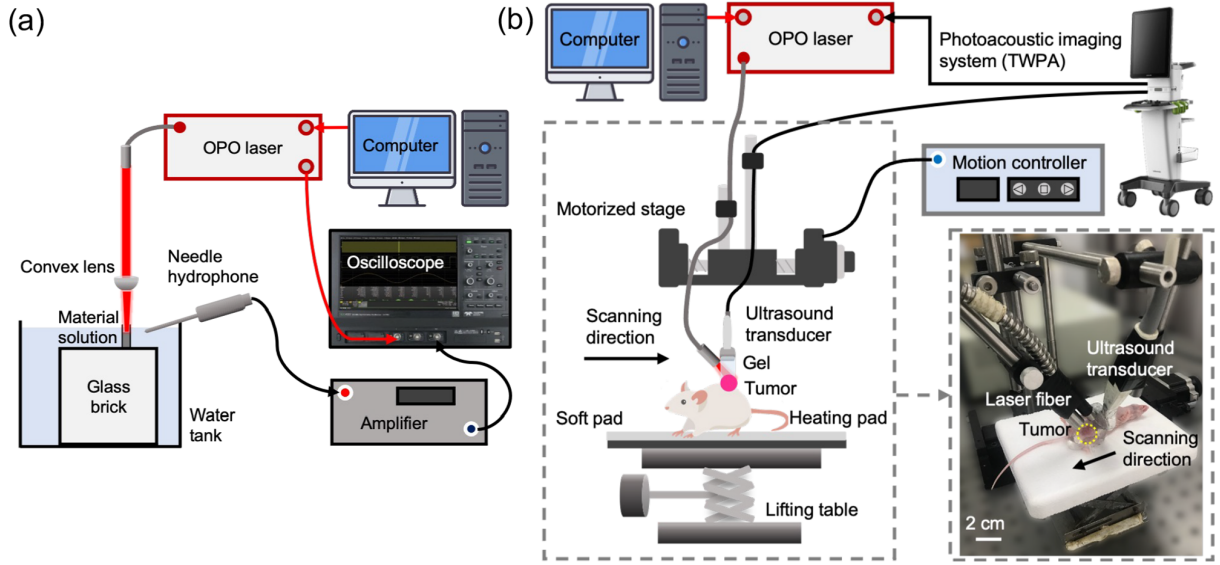

**Figure 1.** Photoacoustic (PA) experimental setup. (a) Experimental system for the in vitro PA characterization of photo-magnetically powered nanoparticles (PMNs). (b) Experimental system for the in vivo 3D photoacoustic imaging (PAI) of intratumoral PMN distribution.

*2.2. Finite element modeling (FEM) simulation of PMNs in bio-environment*

*2.2.1. Electric field.* The electric field induced by the interaction between the laser and PMNs was simulated using the RF module of COMSOL Multiphysics 5.6. The structure of the PMN was modeled using the data obtained from SEM images. The placement of PMNs was set to mimic the assembly of PMNs within the cells under the administration of RMF (260 mT, 10Hz) or RMF+laser (0.5 W/cm$^2$, 1064 nm). The laser (1064 nm, 0.5 W/cm$^2$) was defined as a plane wave that was incident perpendicular to the long axis of the PMN assembly (y direction) and polarized along the axis (x direction). The finite element modeling (FEM) simulation was calculated based on an electromagnetic wave equation:

$$\nabla \times \mu_r^{-1}(\nabla \times \boldsymbol{E}) - k_0^2 \left(\varepsilon_r - \frac{j\sigma}{\omega\varepsilon_0}\right)\boldsymbol{E} = 0 \tag{1}$$

$$\boldsymbol{E}_{laser} = E_0 \cdot \exp(jk_0 y)\hat{x} \tag{2}$$

$$E_0 = \left(\frac{2I}{c\varepsilon_0}\right)^{1/2} \tag{3}$$

Where $\mu_r$ and $\varepsilon_r$ are the relative permeability and dielectric constant. $\boldsymbol{E}$ is the electric field. $k_0 = \omega\sqrt{\varepsilon_0\mu_0}$ is the wavenumber. $\omega$ is the angular frequency. $\mu_0$ and $\varepsilon_0$ are the permeability and the dielectric constant of the vacuum. $\sigma$ is the conductivity. $E_0$ is the electric field amplitude of the incident laser. $I$ is the laser energy density. c is the light speed. Because more than 90% of the PMN was composed of gold whose absorption of light was significantly greater than that of iron, the material of PMN was set as gold in the simulation. The refractive index was referred to the experimental data acquired by Johnson and Christy in 1972 [25]. The physical properties of the PMN and ambient water were obtained from the built-in database pre-stored in COMSOL 5.6.

*2.2.2. Photothermal conversion of PMNs.* The thermal field enhanced by the photothermal conversion of PMNs was calculated using the heat transfer module and the electromagnetic thermal coupling of COMSOL Multiphysics 5.6. The transformation of heat was described as follows:

$$\rho_n C_{p,n} \frac{\partial T_n(\vec{r},t)}{\partial t} = q_e \quad (4)$$

$$\rho_w C_{p,w} \frac{\partial T_w(\vec{r},t)}{\partial t} - k_w \nabla^2 T_w(\vec{r},t) = 0 \quad (5)$$

$$q_e = Re(\boldsymbol{J} \cdot \boldsymbol{E})/2 \quad (6)$$

$$k_w = C_{p,w} \cdot (\rho v_s^2 C_{v,w})^{-1} \quad (7)$$

Where $\rho$ is the density. $C_p$ and $C_v$ are the heat capacity of constant pressure and constant volume. $T$ is the temperature. $q_e$ is the power density of electromagnetic power loss on the PMNs, which is the heat source generated on the PMNs under laser stimulation. $k$ is the thermal conductivity. $\boldsymbol{J} = \sigma \boldsymbol{E}$ is the current density in the PMNs. $v_s$ is the sound speed in the water. $n$, $w$ stands for nanoparticles (NPs) and water, respectively. The temperature at the external boundary of the simulation was kept at a constant of 20°C.

*2.3. In vivo 3D PA imaging (PAI) of PMN distribution in tumors.*

Animal tumor models were established by implanting human triple-negative breast cancer cells (MDA-MB-231) into the groins of female BALB/c athymic nude mice. All nude mice used in this study were raised in the specific pathogen-free animal house of Tongji University, obeying the rules of the institutional animal care and use committee.

PA detection was conducted after the nude mouse tumors reached a diameter of 1 cm (~4 weeks). The nude mice were divided randomly into 3 groups with 3 mice per group: Control, Laser (1064 nm, 0.5 W/cm$^2$, 10 mins), and Laser + rotating magnetic field (RMF) (260 mT, 10 Hz, 10 mins). PMN solution (20 μL, 20 mg/mL) was intratumorally injected into the center of the tumor tissues. For a continuous tracking of the PMN locomotion, we chose 3 time points for PA imaging: before, immediately post, and 24-hour post-injection. Figure 1(b) shows the setup utilized for the 3D scan of mouse tumors. A custom-built ultrasound and PA dual-modality imaging device (TWPA, Tongji University, Shanghai, China & Wisonic, Guangdong, China) was applied with its laser fiber and ultrasound transducer being bound to a motorized linear stage (PA050, Zolix, Beijing, China). The fiber inclined 45 degrees to obtain the best light diffusion within tumors. A motion controller (SC300-3A, Zolix, Beijing, China) changed the stage motion to lead the fiber and transducer to move along the maximum diameters of tumors. A laser of 808 nm (32 mJ/cm$^2$) was exploited for the in vivo PA measurements of this study for the following reasons: First, 808 nm was a frequently used wavelength in the NIR window for biomedical investigation. Second, we were uncertain whether the upper energy limit of our OPO laser at 1200 nm (20 mJ/cm$^2$) was sufficient to penetrate the whole mouse tumor and could produce enough contrast of PMNs. During PA detection, the mice were anesthetized using isoflurane (15%) and placed on a lifting table. The imaging was about 40 seconds, and 1 short video was recorded for each test.

*2.4. In vitro real-time PAI of PMN locomotion driven by the magnetic field (MF).*

Figure 2(a) displays the configuration of devices constructed for the in vitro PA imaging (PAI) of PMNs manipulated by the MF. The ultrasound transducer (L22-10-H, Wisonic, Guangdong, China) and the self-designed laser outlet were held together by a 3D-printed fixture. A glass tube containing the PMN solution (0.5 mg/mL) was placed below the transducer. The magnet was moved at a low speed to ensure that PMNs in the tube closely followed its motion. The laser outlet had a square shape with evenly distributed fibers. As shown in Figure 2(b), the laser outgoing from the outlet was reflected once by a mirror stuck on the fixture and irradiated at a 45-degree angle on the tube. The connection of the ultrasound transducer and laser fiber to other equipment was the same as depicted in Figure 1(b).

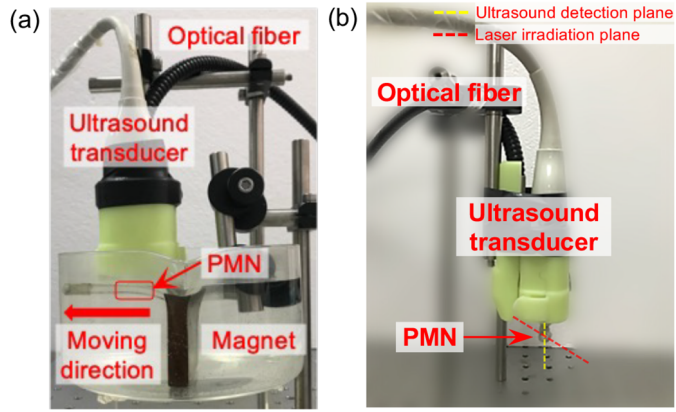

**Figure 2.** Configuration of devices constructed for the magnetic actuation and in vitro PAI of the photo-magnetically powered nanoparticle (PMN). (a) Front view. (b) Side view.

### 2.5. Data processing methods.

The data processing of this study was accomplished in MATLAB 2021a. The PA signal amplitude was obtained from the peak-to-peak value for the in vitro PA measurements of PMNs (section 2.1.2).

For the in vivo 3D PA scanning of nude mouse tumors (section 2.3), the frames of videos were extracted and segmented to obtain the ultrasound and PA parts separately. The distribution of PMNs within the tumors was quantified according to the following equation [22]:

$$Distribution = \frac{count(I_{pixel} > I_{threshold})|ROI}{V_{cancer}} \quad (8)$$

$$V_{cancer} = 0.5 \times D_{max} \times D_{min}^2 \quad (9)$$

Where $count$ stands for the number of pixels. $I_{pixel}$ and $I_{threshold}$ are the intensity of the pixel and the threshold for PMNs in PA images. $I_{threshold}$ was defined as 40% of the maximal pixel intensity. $ROI$ is the region of interest: the inner area of cancer, which is demarcated based on the ultrasound images. $V_{cancer}$ is the volume of the cancer. $D_{max}$ and $D_{min}$ are the maximal and minimum diameters of the cancer. For a clear visualization of PMN distribution in each mouse xenograft, the ultrasound and PA images were fused and reconstructed into 3D volume-rendered images using FUJI ImageJ software.

For in vitro PAI of PMN locomotion (section 2.4), the video frames were first segmented and fused as described above. Then the images were reprinted in pseudo-color and fabricated into a video.

### 2.6. Statistical analysis

The statistical analysis in this study was accomplished using GraphPad Prism 9.0. Data were provided as mean ± standard deviations (SDs) for 3 samples per group. The correlation between the PA signal amplitude and the PMN solution concentration was studied via simple linear regression.

## 3. Results

### 3.1. Basic properties of PMNs

As shown in Figure 3(a), the PMNs designed in this study owned uniform morphology with a core and a spiky shell. The statistical results obtained from the SEM and TEM images displayed that PMNs had an average diameter of 509 ± 38 nm, with their tips owning a length of 128 ± 20 nm and a 24 ± 4 nm diameter at the bottoms. Gold occupied a crucial role in the composition of PMNs (more than 90%), which acted as a strong light absorber that produced PA contrast. Iron was a small fraction of

PMNs, which ensured the MF control of PMNs. Figure 3(b) exhibits the UV-vis to NIR extinction spectrum of PMNs; it represented a broad absorbance peak in the NIR-I and NIR-II regions. The PA signals of PMNs increased linearly with the concentrations. PMNs remained stable within the illumination of a pulsed laser (1200 nm, 10 Hz, 20 mJ/cm$^2$) for 60 seconds, as depicted in Figure 3(c) and 3(d). These results revealed a great potential for applying PMNs and PAI tracking in cancer mechanotherapy in vivo.

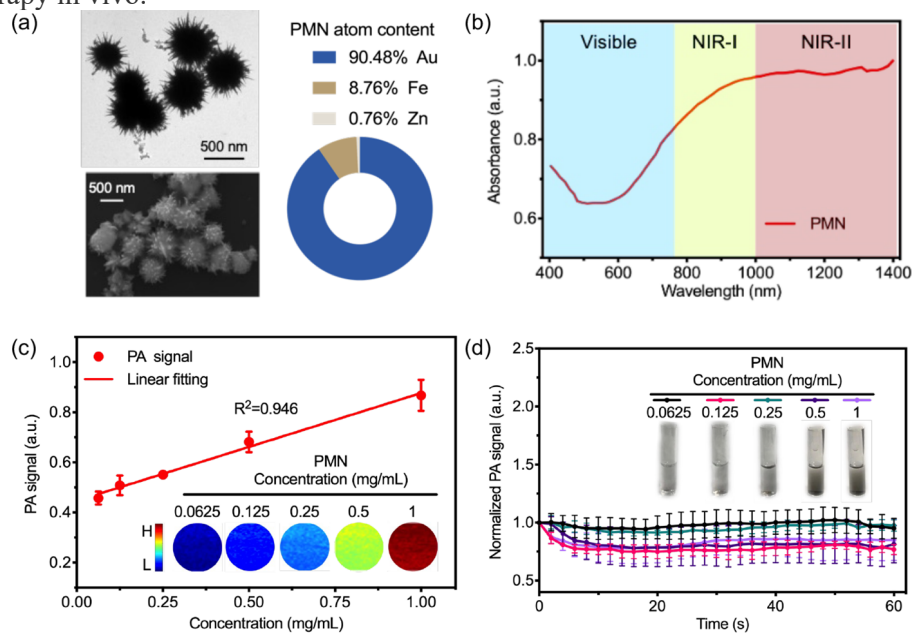

**Figure 3.** Characterization of the PMNs. (a) Representative transmission electron microscopy (TEM) and scanning electron microscopy (SEM) of PMNs with the contents of metal atoms. (b) Vis-NIR spectrum of PMNs. (c) PA signals of PMNs at 808 nm as a function of concentration (mean ± s.d, n=3), inset: PA images. (d) Normalized PA signal of PMNs under a pulsed laser illumination (1200 nm, 10 Hz, 20 mJ/cm$^2$) for 60 seconds. Inset: photographs of PMN solution of various concentrations.

### 3.2. MF-triggered locomotion of PMNs

Figure 4 shows the real-time PAI of PMN movements led by the external MF. The captured frames proved that MF could manipulate the PMNs efficiently, and the real-time position of PMNs could be visualized via PAI. The laser fluence on the left side of the fiber outlet was significantly larger than that on the right side. The results indicated that the contrast of PAI could be tuned by adjusting the laser energy.

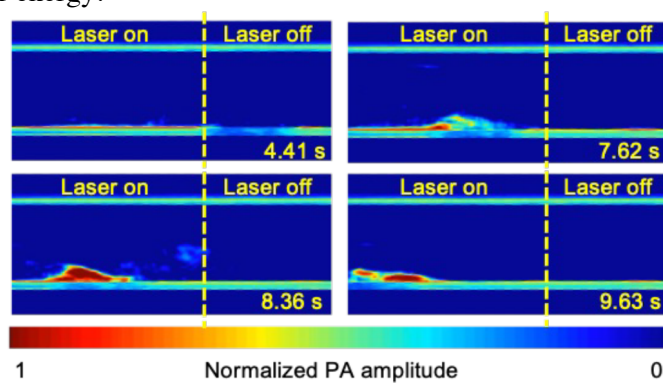

**Figure 4.** Time-lapse PA images of PMN locomotion manipulated by the magnetic field (MF).

*3.3. Assembly of PMNs in cells under external stimuli*

The Bio-TEM observation results in Figure 5(a) exhibited that PMNs assembled into short chains in cells under the administration of an RMF (260 mT, 10 Hz) for 5 mins. Figure 5(b) shows that elongated PMN chains were formed when the RMF and laser (0.5 W/cm$^2$, 1064 nm) were applied simultaneously. Simulations of the electric and thermal field were conducted to discover the enhanced light interaction contributed by the PMN assembly on the microscale. From the comparison of Figure 5(c) and Figure 5(d), we discovered that the elongated chain could induce a stronger electric field. Consequently, the photothermal conversion via PMNs was increased under the dual external stimuli, as shown in Figures 5(e) and Figure 5(f).

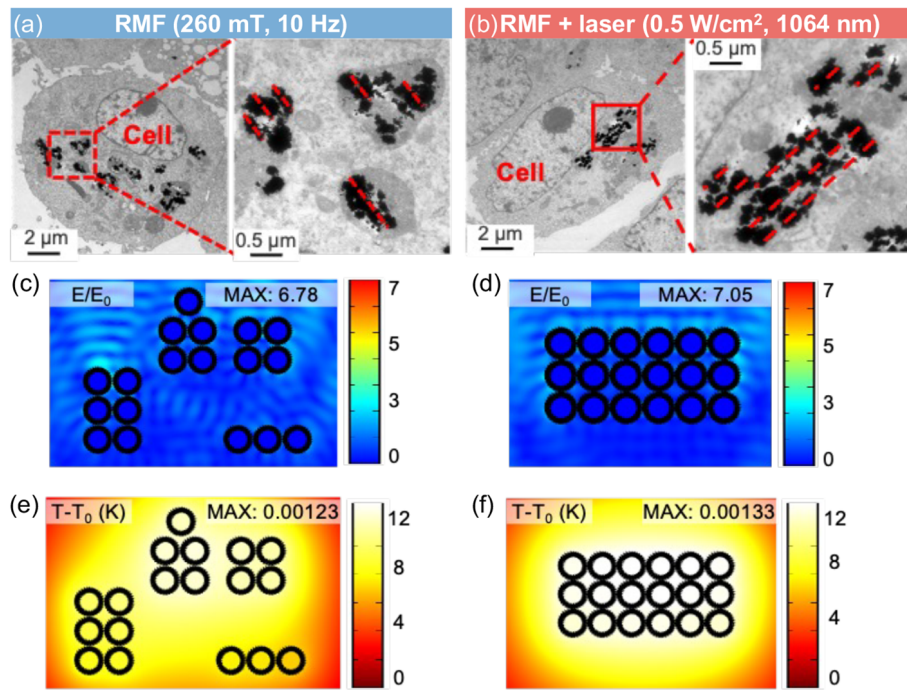

**Figure 5.** The assembly of PMNs in cells. Bio-TEM image of PMNs in cells subjected to (a) a rotating magnetic field (RMF) (260 mT, 10 Hz) and (b) RMF with a continuous wave (CW) laser (0.5 W/cm$^2$, 1064 nm) for 5 mins. Electric field simulations of PMNs treated with (c) only RMF and (d) RMF with laser. Thermal field simulations of PMNs treated with (e) only RMF and (s) RMF with laser.

*3.4. Longitudinal monitoring of PMN distribution in tumors*

Figure 6(a)-(c) represents the continuous PA tracking results of intratumoral PMN distribution. The quantified results were shown in Figure 6(d) according to equation (8) and equation (9). The lifted PMN distribution of the laser group at 0 h was because of the extended injection position. However, the distribution soon reduced to the baseline due to the high tumor interstitial pressure [26]. Tumors subjected to the CW laser (0.5 W/cm$^2$, 1064 nm) and RMF (260 mT, 10 Hz) for 10 mins had greater increases in PMN distribution after 24-hour injection compared to those of the control and the laser groups.

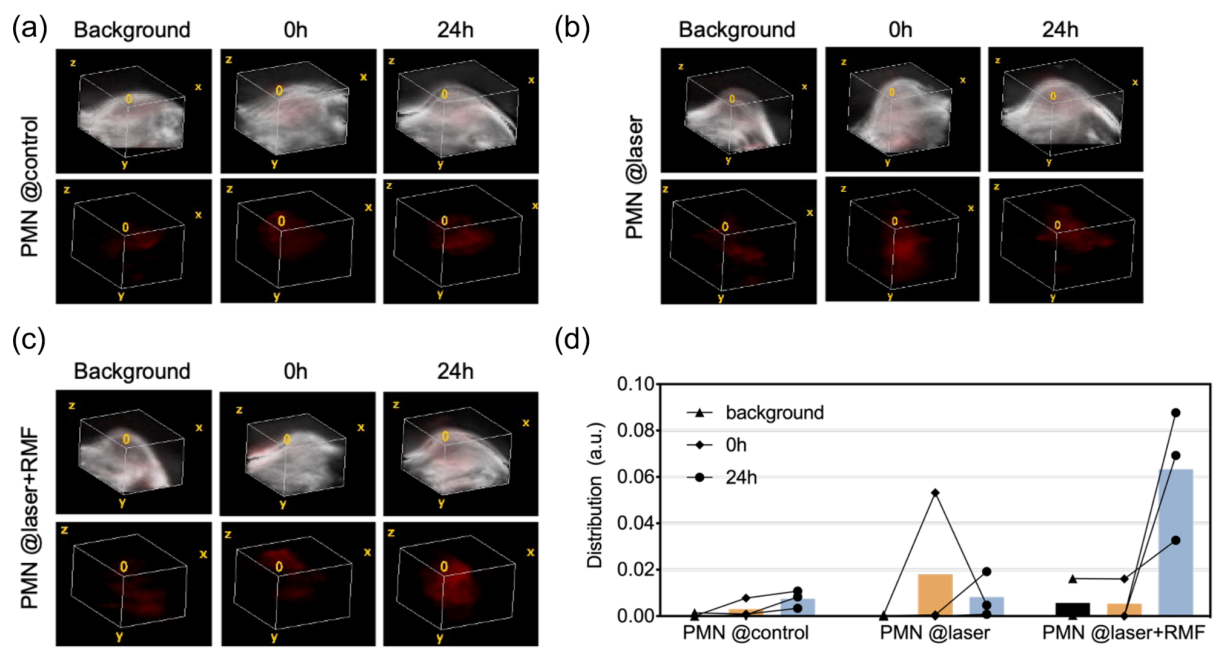

**Figure 6.** Intratumoral PMN distribution under external stimuli. 3D PAI of PMN distribution in tumors (a) without external stimuli and under the stimuli of (b) CW laser (0.5 W/cm$^2$, 1064 nm, 10 mins) or (c) CW laser with RMF (260 mT, 10 Hz, 10 mins). (d) Quantitative results of three different experimental conditions.

## 4. Discussion

The study of functional NPs is at an expanding stage with the growing demands of precision medicine. The manipulation of NPs provides an efficient technique to destroy the pathological tissues mechanically, which is feasible for various lesions [27]. The PA method has the advantages of non-invasiveness, non-radiation, and high penetration depth, which is quite suitable for the in vivo tracking of NPs. Therefore, the application of NPs with PAI tracking shows excellent potential for the clinical transformation of cancer therapy.

This study developed a novel NP with a Fe-Au core-shell structure. This PMN can respond simultaneously to the laser and MF as verified by the PA detection and ex vivo real-time PAI. The gold nano-tip-covered shell of the PMN has two critical functions that benefit cancer therapy: Firstly, the destructive pressure exerted on the ambient tissues is enhanced [7]. Secondly, the photothermal conversion is increased, which was proved by the simulation works; this can reduce the bio-environment viscosity [28]. PMNs in cells subjected to the laser and RMF assembled into elongated chains compared to those under only RMF as shown by the bio-TEM observations; this suggested that the in vivo moving ability of PMNs was improved. Then, the continuous 3D PAI of nude mouse xenografts was performed to quantify the PMN distribution. The intratumoral volume proportion of PMNs under the administration of dual external stimuli was greater than that of no stimuli or only laser; it indicated that the coupling of laser and MF could enhance the locomotion and retention of PMNs within tumors, which offered a feasible strategy to benefit cancer therapy.

Our research is still incomplete, and more effort needs to be made in our future works. First, we contrive to design different MFs to precisely control PMNs; this can further minimize the invasiveness and improve the effect of cancer mechanotherapy. Second, we plan to use PAI to monitor the dynamic process of PMN assembling. Third, we intend to utilize PAI to analyze the rotational motion of PMN chains under RMF. These two works aim to determine whether the actuated movement of PMNs and their assembly can reflect the physical properties of the bio-environment and the destructive power exerted on the cancer tissues.

## 5. Conclusion

This study demonstrated the feasibility of utilizing PAI to monitor the locomotion of PMNs propelled by the laser and MF intended for cancer therapy. We synthesized an NP with an iron oxide core and a gold-nanotip-covered shell, which has strong light absorption in the NIR-II region because of the enhanced LSPR effect. The devised PMN is sensitive to the MF, as the ex vivo real-time PAI proves, and could produce strong and stable PA signals. The laser and MF coupling effect can significantly enhance energy conversion via PMNs. According to the Bio-TEM observation, elongated PMN chains were formed in cells subjected to dual external stimuli. As explained by the simulations, photothermal conversion was increased when more PMNs were activated to form a longer assembly; this can further reduce the viscosity of the bio-environment. The continuous 3D PAI results elaborated that the laser and MF coupling can elevate the PMN distribution and retention in mouse tumors after injection. The above results confirmed that PMN stimulated by laser-MF coupling with PAI monitoring provides a novel platform for precise cancer mechanotherapy with high efficacy.

**Acknowledgments**

This project was supported by the National Natural Science Foundation of China (grant number 12034015), the Program of Shanghai Academic Research Leader (grant number 21XD1403600), and the Shanghai Municipal Science and Technology Major Project (grant number 2021SHZDZX0100).